\documentclass[12pt]{iopart}
\usepackage{iopams}
\usepackage{graphicx}
\expandafter\let\csname equation*\endcsname\relax
\expandafter\let\csname endequation*\endcsname\relax
\usepackage{amsmath,amsfonts,amssymb}

\begin{document}

\title[Competing synchronization on random networks]{Competing synchronization on random networks}

\author{Jinha Park and B. Kahng$^\dagger$}
\address{CCSS, CTP and Department of Physics and Astronomy, Seoul National University, Seoul 08826, Korea}
\ead{$^\dagger$bkahng@snu.ac.kr}

\vspace{10pt}
\begin{indented}
\item[]\today
\end{indented}

\begin{abstract}
The synchronization pattern of a fully connected competing Kuramoto model with a uniform intrinsic frequency distribution $g(\omega)$ was recently considered. This competing Kuramoto model assigns two coupling constants with opposite signs, $K_1 < 0$ and $K_2 > 0$, to the $1-p$ and $p$ fractions of nodes, respectively. This model has a rich phase diagram that includes incoherent, $\pi$, and traveling wave (TW) phases and a hybrid phase transition with abnormal properties that occurs through an intermediate metastable $\pi$ state. Here, we consider the competing Kuramoto model on Erd\H{o}s--R\'enyi (ER) random networks. Numerical simulations and the mean-field solution based on the annealed network approximation reveal that in this case, when the mean degree of the random networks is large, the features of the phase diagram and transition types are consistent overall with those on completely connected networks. However, when the mean degree is small, the mean-field solution is not consistent with the numerical simulation results; specifically, the TW state does not occur, and thus the phase diagram is changed, owing to the strong heterogeneity of the local environment. By contrast, for the original Kuramoto oscillators, the annealed mean-field solution is consistent with the numerical simulation result for ER networks.
\end{abstract}

\vspace{2pc}
\noindent{\it Keywords}: competing Kuramoto model, synchronization, random network

\section{Introduction}

The Kuramoto model (KM)~\cite{kuramoto} is a coupled phase oscillator model that exhibits a synchronization transition from an incoherent (IC) to a synchronized state. This model is written as
\begin{equation}\label{KM}
\dot\theta_i = \omega_i + \frac{K}{N} \sum_{j=1}^{N} \sin(\theta_j-\theta_i),
\end{equation}
where $\theta_i$ denotes the phase of oscillator $i$; $\omega_i$ is the natural frequency of oscillator $i$, which follows the distribution $g(\omega)$; $K$ is a positive coupling constant; and $N$ is the number of oscillators in the system. We note that each oscillator interacts with all the other oscillators, which is characteristic of a conventional mean-field model. This KM has been used as a prototypical model to understand synchronization phenomena in diverse real-world systems ranging from biological systems such as fireflies flashing to physical systems such as electric power grids~\cite{book1,book2,book3}. In the original KM, $g(\omega)$ has a Gaussian distribution, and a continuous synchronization transition occurs as the coupling strength $K$ is increased.  

The coupling constant is not necessarily positive but can be generalized to a mixture of positive and negative constants when the system contains competing interactions. Here, we consider a KM with competing coupling constants, which is inspired by neural networks with excitatory and inhibitory couplings between neurons, and by spin glass systems with ferromagnetic and antiferromagnetic interactions. For instance, such competing interactions in spin glass systems often result in the frustration of magnetic ordering and the emergence of a glass state. Metastable slow dynamics proceeds across the free energy landscape. Therefore, similar metastable behaviors may be expected to arise in the competing KM ($c$-KM).

The KM with competing interaction was first considered by Daido~\cite{daido} in 1992. He used the Sherrington--Kirkpatrick random coupling constant $K_{ij}$ between all pairs of oscillators $(i,j)$, which follows a Gaussian distribution with zero mean and a finite standard deviation. Under this type of competing interaction, several interesting behaviors appeared, for instance, quasi-entrainment, diffusive dynamics in phase space, and a slow power-law decay of coherence. Furthermore, the possible existence of a glassy oscillating state, which leads to the so-called volcano transition, was suggested~\cite{daido, volcano}.  

Further, Hong and Strogatz~\cite{Hong11id,Hong11,Hong12} considered a tractable KM with node-based random couplings $\{K_i\}$, in which a $1-p$ fraction of the nodes have a negative coupling constant $K_1 <0$, and the remaining nodes have a positive coupling constant $K_2$. The authors adopted a Lorentzian $g(\omega)$ and used the Ott--Antonsen (OA) ansatz~\cite{Ott08,Ott09}; they found that the synchronized state of this $c$-KM can be characterized by the dynamics of two groups of oscillators roughly separated by an angle $\pi$ in phase space. They are either static or traveling. 

The OA method has been successful in the bifurcation analysis of several KMs with unimodal $g(\omega)$ such as a Lorentzian distribution~\cite{Hong11,Hong12,Ott08,Ott09,Martens09,vanHemmen}. The reason is that an $N\rightarrow\infty$ continuum of the oscillator degrees of freedom is exactly reduced to a few coupled modes. The Watanabe--Strogatz method is also useful for reducing the degrees of freedom of the coupled nonlinear equations of $N$ identical oscillators~\cite{Hong11id,Watanabe,moebius,Pikovsky}. The order parameter dynamics is exactly calculable in these OA and Watanabe--Strogatz reducible systems. However, for a uniform distribution, reduction is hardly achieved. Instead, one can at best seek the stationary solution using the complex self-consistency (SC) equation. 

In our previous study~\cite{cpazo_pre}, we considered a $c$-KM model with a uniform $g(\omega)$ in the interval $[-\gamma,\gamma]$, which is called the competing $c$-Winfree--Paz\'o ($c$-WP) model. Using the complex SC equation, we showed briefly that a discontinuous synchronization transition with critical behavior, which is often called a hybrid phase transition, occurs. We focused mainly on the two-step discontinuous transitions that arise above the hybrid transition point: the transitions from the IC to the $\pi$ state, and then to the traveling wave (TW) state, where the intermediate $\pi$ state is characterized by a long period of metastability. 

In this study, we first recapitulate a general formalism of the complex SC equation that is applicable to general types of $g(\omega,K)$~\cite{cpazo_pre,stefanovska}. Next, we derive explicitly the complex SC equation for the $c$-WP model. In this case, the $\pi$ and TW states can emerge in the synchronized regime. The imaginary part of the SC equation determines the TW solution. A rich phase diagram is obtained, and critical behaviors are derived explicitly as a function of the parameter set $\{p, Q\equiv |K_1|/K_2, \gamma\}$. However, the complex SC equation does not determine the stability of the solution rigorously. Instead, the empirical linear stability of each complex SC equation solution can be analyzed~\cite{stefanovska}. However, the linear stability did not distinguish the metastable states that arise in the $c$-WP model~\cite{cpazo_pre}. 

Next, we consider the $c$-KM on Erd\H{o}s--R\'enyi (ER) networks. First, using the mean-field theory for annealed ER networks, we check analytically whether the synchronization transitions of the all-to-all coupling case occur even on ER networks. We find that the synchronization transitions are similar overall. However, the simulation results show significant differences. On ER networks, the mean degree of the ER network, $\langle q \rangle$, also plays an important role in determining the synchronization transition properties. When $\langle q \rangle$ is small, the TW state can disappear, even though the analytic and numerical solutions for annealed ER networks predict its existence. Moreover, the transition type can change.

This paper is organized as follows. In Sec.~\ref{sec:model}, we introduce the $c$-KM and investigate the origin of the TW state of the oscillators. The mean-field solution of the TW is also presented. In Sec.~\ref{sec:csc}, we derive the complex SC equation using the original Kuramoto method and the OA ansatz. In Sec.~\ref{sec:wp_model}, we consider the $c$-WP model and derive the SC equation explicitly. Using the SC equation, we derive the order parameters $R$ and $\Omega$ of the $c$-WP model for various $Q=|K_1|/K_2$ and the half-width $\gamma$ of the uniformly distributed $g(\omega)$. We obtain rich phase diagrams including IC, $\pi$, and TW phases and continuous, discontinuous, and hybrid phase transitions. In Sec.~\ref{sec:er}, we consider the $c$-KM model on ER networks. Using the mean-field approach, we obtain the mean-field SC solution for annealed ER networks and the corresponding phase diagrams. Both annealed and quenched ER networks are numerically simulated. The order parameter curves are obtained as a function of $p$ for various values of $\gamma$, $Q$, and the mean degree $\langle q \rangle$. We observe some discrepancies between the mean-field solution and the quenched ER network results, and consider possible reasons for the inconsistencies. In Sec.~\ref{sec:summary}, we summarize our results.

\section{The competing Kuramoto model}\label{sec:model}
The $c$-KM on completely connected networks is written as 
\begin{equation}
\dot{\theta_i} = \omega_i + \frac{K_i}{N}\sum_{j=1}^{N} \sin(\theta_j-\theta_i), \quad i=1,2,\cdots,N,
\label{eq:km}
\end{equation}
where the phase $\theta_i$ of each oscillator has an intrinsic frequency $\omega_i$, and $\omega_i$ follows the distribution function $g(\omega)$. In this section, we do not assume a specific form of $g(\omega)$. $K_i$ is the coupling constant of oscillator $i$. A fraction $(1-p)$ of the nodes have a coupling constant $K_1 < 0$, and the remaining nodes have $K_2 > 0$. In addition, $p$ is a control parameter. This $c$-KM can be generalized to an arbitrary number of species of coupling constants. 

The synchronization transition is characterized by a complex order parameter at time $t$,
\begin{equation}
Z(t)\equiv R(t)e^{i\psi(t)}\equiv\frac{1}{N}\sum_{j=1}^{N}e^{i\theta_j(t)},
\label{eq:c_op}
\end{equation}
where $R$ is the coherence of the oscillators, and $\psi$ is the average phase. In the long-time limit, the system falls into the stationary state, in which $Z\approx Re^{i\Omega t}$. The synchronization order parameter $R$ distinguishes the coherent (C) and IC phases. The C phase is further divided into the $\pi$ and TW phases by the TW order parameter $\Omega$. Overall, three phases are possible: the IC ($R=0$), $\pi$ ($R\neq0,\Omega=0$), and TW ($R\neq0,\Omega\neq0$) phases. 

\begin{figure}[!ht]
	\centering
	\includegraphics[width=0.7\linewidth]{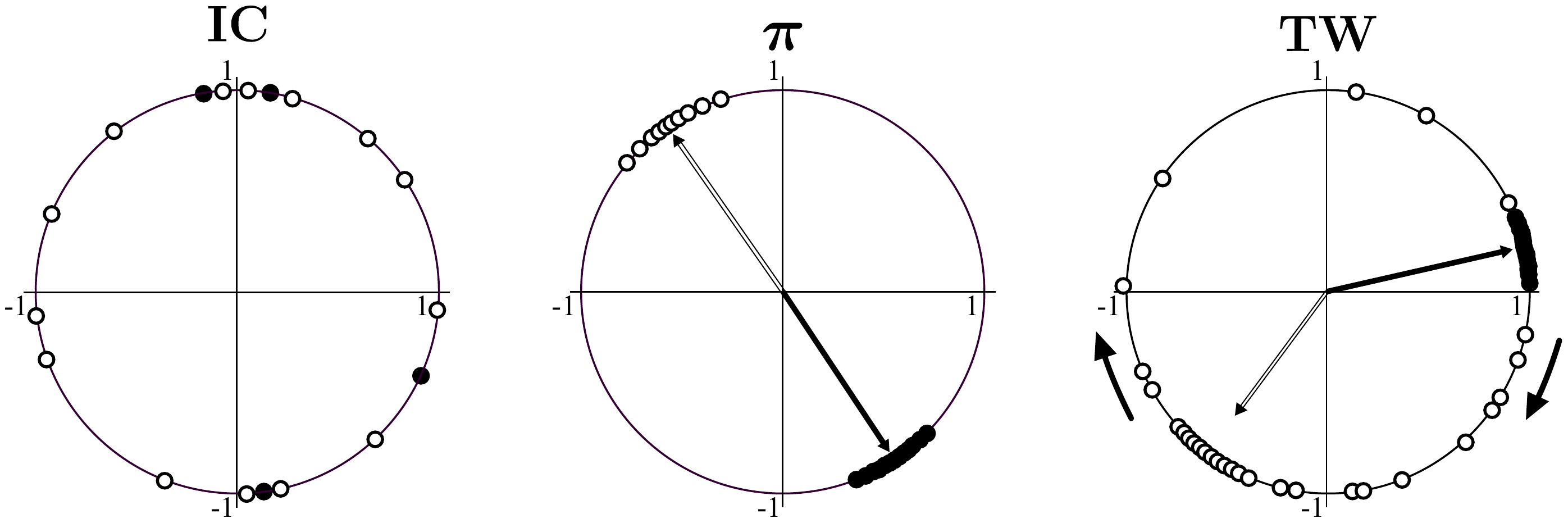}
	\caption{Schematic illustration of the IC, $\pi$, and TW states. Depending on the sign of the coupling constant $K_i$, the stability of oscillator $i$ at a velocity-balancing position is reversed. Each oscillator with  $K_2>0$ ($\bullet$) is attracted toward the mean phase $\psi$, whereas each oscillator with $K_1 < 0$ ($\circ$) is drawn towards the antipode $\psi+\pi$. Therefore, the oscillators may separate into two groups according to the signs of their coupling constants. Two types of steady C states are possible. The two groups can be separated by an angle $\pi$ and remain static; otherwise, they can rotate at a common TW speed $\Omega$.}
	\label{fig:group-sync}
\end{figure}

Inserting the definition of the complex order parameter into \eqref{eq:km} results:
\begin{equation}
\dot{\theta_i} = \omega_i + K_i R(t)\sin\left(\psi(t)-\theta_i\right), \quad i=1,2,\cdots,N.
\label{eq:km_op}
\end{equation}
Depending on the sign of the coupling constant $K_i$, the stability of an oscillator at a velocity-balancing position is reversed. Thus, the oscillators are tiered into two groups on the phase circle. Each oscillator with $K_2>0$ is attracted toward the average phase $\psi$ with a strength proportional to the coherence $R$, and each oscillator with $K_1<0$ is drawn toward the antipode $\psi+\pi$, as shown in the schematic illustration in Fig.~\ref{fig:group-sync}. Therefore, in the synchronized phase, the oscillators are clustered into two groups, where each positive and negative coupling population forms a separate group. The $K_1$ group, i.e., the group of oscillators with the coupling constant $K_1 < 0$, generally shows a broader angular distribution than the $K_2$ group, because each $K_1$ oscillator repels all the other oscillators, whereas each $K_2$ oscillator attracts all the other oscillators. 

In the $\pi$ phase, the two groups are balanced at a separation angle $\pi$ and occupy a constant position on the phase circle. In the TW phase, a separation of less than $\pi$ is maintained, where the attractive $K_2$ group tries to catch up with the $K_1$ group, and the repulsive $K_1$ group tries to maintain its distance from the $K_2$ group. As a result, the two groups co-rotate around the phase circle at a common angular speed $\Omega$.

\subsection{The origin of the TW phase}
This non-static steady state is a novel property of $c$-KMs~\cite{Hong11}. The TW phase is possible in the competing model owing to the asymmetric interactions. In particular, competing interactions may cause a nontrivial net rotation $\Omega$ in the system. Note that $\Omega$ is different from the average natural frequency, which is zero.

The TW phase can emerge only if the attractive force of the $K_2$ group is sufficiently larger than the repulsive force of the $K_1$ group, i.e., if $Q\equiv |K_1|/K_2 < 1$. Otherwise, a separation of less than $\pi$ is unstable. We construct a mean-field theory below as a first attempt to understand the emergence of the TW phase.

\subsection{Mean-field theory for the TW phase}
In the TW phase, the system rotates collectively at a nonzero speed. The average angular speed~\cite{stefanovska_tw} of the groups $K_1$ and $K_2$ and that of the total system are calculated as follows:
\begin{eqnarray}
v_1 &\equiv \langle \dot\theta_j\rangle_1 \equiv \frac{1}{N_1}\sum_{j\in K_1}\left[\omega_j +\frac{K_1}{N}\sum_{k=1}^{N} \sin(\theta_k-\theta_j)\right] 
= p |K_1| r_1 r_2 \sin\Delta, \nonumber \\\vspace{5mm}
v_2 &\equiv \langle \dot\theta_j\rangle_2 \equiv \frac{1}{N_2}\sum_{j\in K_2}\left[\omega_j +\frac{K_2}{N}\sum_{k=1}^{N} \sin(\theta_k-\theta_j)\right]
= (1-p) K_2 r_1 r_2 \sin\Delta, \nonumber\\ \vspace{5mm}
v &= p v_2 + (1-p) v_1 
= p(1-p)(|K_1|+K_2)r_1r_2\sin\Delta,
\end{eqnarray}
where $\langle e^{i\theta_j}\rangle_{1,2}\equiv r_{1,2}e^{i\psi_{1,2}}$, and the separation between the two groups is denoted as $\psi_1-\psi_2\equiv\Delta$. For $g(\omega)$ with even symmetry, $\langle\omega_j\rangle_1=\langle\omega_j\rangle_2=0$. The angle brackets denote averaging over a group of oscillators. In addition, $\langle\sin\theta_{12}\rangle \equiv \frac{1}{N_1N_2}\sum_{j\in K_1}\sum_{k\in K_2}\sin(\theta_j-\theta_k)= r_1 r_2 \sin(\psi_1-\psi_2) = r_1r_2\sin\Delta$. The summation counts only for intergroup interactions because the intragroup interactions cancel out pairwise to zero. When $\Delta=\pi$, $\sin\Delta$ is zero, and therefore $v_1=v_2=v=0$. However, in the TW phase, $\Delta$ is less than $\pi$, and the angular speed $v$ is nonzero. Now for each group $\alpha=1$ and $2$,
\begin{equation}
\psi_\alpha = \arctan \left[\frac{\langle\sin\theta_i\rangle_\alpha}{\langle\cos\theta_i\rangle_\alpha}\right] 
=\arctan \left[ \frac{\langle\theta_i\rangle_\alpha -\langle\theta_i^3\rangle_\alpha/3! +\cdots}{1 -\langle\theta_i^2\rangle_\alpha/2! +\cdots} \right].
\end{equation}
As a mean-field approximation, we let $\Delta\approx\langle\theta_{12}\rangle \equiv \frac{1}{N_1N_2}\sum_{j\in 1}\sum_{k\in 2}(\theta_j-\theta_k)$ and $\psi_\alpha\approx\langle\theta_i\rangle_\alpha(1+(\langle\theta_i^2\rangle_\alpha-\langle\theta_i\rangle_\alpha^2)/2!)\approx\langle\theta_i\rangle_\alpha$ 
for each group; i.e., $r_1\approx r_2\approx 1$. The approximation is valid as long as the distribution angle of each group is small. Let $\Delta\equiv\pi-\delta$. Assuming that the amplitude components are stable, the phase dynamics of $\delta$ is given as 
\begin{equation}
\dot\delta \approx - \langle \dot\theta_{12}\rangle = -(v_1-v_2) = - (p-p_u) (|K_1|+K_2)r_1r_2\sin\delta,
\end{equation}
where
\begin{equation}
p_u=\frac{1}{Q+1}.
\end{equation}
Note that for $p>p_u$, the $\pi$ state ($\delta=0$) is a stable solution. The $\pi$ state becomes unstable at $p=p_u$ as $p$ is decreased, suggesting that a new solution with $\Delta\neq\pi$ (the TW phase) may emerge for $p < p_u$. Using the SC equation, we find that the TW solution indeed exists in some interval $[p_\ell,p_u]$. However, this lower bound $p_\ell$ is not determined from the mean-field calculation.

\section{The complex SC equation}\label{sec:csc}  
Here, we search for steady-state solutions $Z\approx R e^{i\Omega t}$. For this purpose, we apply the SC method as follows. In the rotating frame $\phi_j \equiv \theta_j -\Omega t$, each oscillator follows
\begin{equation}
\dot\phi_j = \omega_j-\Omega- K_j R \sin\phi_j. 
\end{equation}
Depending on the relative magnitudes of the frequency disorder $\omega_i-\Omega$ and the coupling strength $|K_j R|$, each oscillator is either phase-locked or drifting. The order parameter is written as 
\begin{equation}
R = \frac{1}{N}\sum_j \overline{e^{i\phi_j(t)}},
\end{equation}
and the time average, which is denoted by the overline, is taken in the steady-state regime for self-consistency. Each oscillator with a natural frequency satisfying $|\omega_j-\Omega|\leq|K_j|R$ becomes phase-locked at
\begin{equation}
\phi_j^*= \arcsin\left[\frac{\omega_j-\Omega}{K_j R}\right],
\label{eq:entrained}
\end{equation}
which ultimately contributes to the total synchronization order. In addition, each drifting oscillator satisfying $|\omega_j-\Omega|\geq |K_j|R$ will repeatedly move ahead of or fall behind the locked population. Their rotation period is calculated as 
\begin{equation}
T_j = \int_{0}^{2\pi}\frac{d\phi_j}{|\dot\phi_j|}= \frac{2\pi}{\sqrt{(\omega_j-\Omega)^2-(K_jR)^2}}.
\end{equation}
Therefore, the contribution of the drifting oscillators to the order parameter $R$ is zero. Indeed, individual drifting oscillators contribute only to the order parameter $\Omega$ as 
\begin{eqnarray}	
\overline{e^{{\rm i}\phi_j}} &= \frac{1}{T_j}\int_0^{2\pi} \frac{d\phi_j}{\dot\phi_j} e^{{\rm i}\phi_j} \nonumber \\
&=\frac{{\rm i}~\textrm{sgn}(\omega_j-\Omega)}{K_j R} \left[|\omega_j-\Omega|-\sqrt{(\omega_j-\Omega)^2-(K_jR)^2}\right].
\end{eqnarray} 
After some calculations, the complex SC equation is obtained as follows:
\begin{eqnarray}
R &= \int_{-\infty}^{\infty}dK d\omega g(K,\omega)\overline{e^{{\rm i}\phi}} \nonumber \\ 
&=\int_{-\infty}^{\infty}dK \int_{\Omega-|K|R}^{\Omega+|K|R} d\omega g(K,\omega)~\textrm{sgn}(K) \sqrt{1-\left(\frac{\omega-\Omega}{KR}\right)^2} \nonumber \\
&\hspace{1em}+{\rm i} \int_{-\infty}^{\infty} dK \int_{\rm drifting} d\omega \frac{g(K,\omega)}{KR} \Big[\omega-\Omega \nonumber\\ &\hspace{13em}-\textrm{sgn}(\omega-\Omega)\sqrt{(\omega-\Omega)^2-(KR)^2}\Big],
\label{eq:sce}
\end{eqnarray}
where $g(K,\omega)$ is the distribution of disorder in the $c$-KM, and ${\rm sgn}(x)=1$ and $-1$ for $x > 0$ and $x < 0$, respectively. 

The imaginary part on the right-hand side of the complex SC equation \eqref{eq:sce} has to vanish, because $R$ on the left-hand side is real-valued. In the KM, $\Omega$ should generally be given as the mean intrinsic frequency of the system; i.e., the TW order is absent. Therefore, only the real part of the equation must be solved. Indeed, when $g$ is symmetric with respect to $\Omega$, one can easily check that the integrand of the imaginary part becomes odd in $\omega$ and goes to zero. To obtain the TW solution, however, the real and imaginary parts of Eq.~\eqref{eq:sce} must be solved simultaneously for $R$ and $\Omega$.

The complex SC equation can also be obtained by applying the OA ansatz to the continuum version of the KM. 
\begin{equation}
\frac{\partial \rho}{\partial t} + \frac{\partial}{\partial\theta}\left[\rho(K,\omega,\theta,t)\left(\omega+K\frac{Z e^{-i\theta}-Z^*e^{i\theta}}{2i}\right)\right]=0.
\end{equation}
The density function is assumed to have the form  $$\rho=\frac{g(K,\omega)}{2\pi}\left[1+\sum_{n=1}^{\infty}\left(a_{\omega,K}^n(t)e^{in\theta}+ {\rm c.c.}\right)\right],$$ where $a_{\omega,K}$ satisfies the equation
\begin{equation}
\dot{a}_{\omega,K}(t) = -i\omega a_{\omega,K} + \frac{K}{2}(Z^*-Za_{\omega,K}^2).
\label{eq:oa-mode-eq}
\end{equation}
The order parameter is then given as
\begin{eqnarray}
Z(t) &= \int dK d\omega d\theta e^{i\theta}\rho(K,\omega,\theta,t) =\int dK d\omega g(K,\omega) a_{\omega,K}^*(t).
\label{eq:oa-sce}
\end{eqnarray}
In the steady state, Eqs.~\eqref{eq:oa-mode-eq} and \eqref{eq:oa-sce} also lead to the complex SC equation that we derived as Eq.~\eqref{eq:sce}~\cite{stefanovska}. For the Lorentzian frequency distribution $g(\omega)=\frac{\gamma/\pi}{\omega^2+\gamma^2}$, the above integration can be performed in the complex domain, and only a single complex mode, $a^*_{-i\gamma,K}$ (for each $K$), contributes to the order parameter. This dimensional reduction does not generally occur for other types of frequency distributions; however, Eq.~\eqref{eq:oa-sce} remains valid. Therefore, exact bifurcation analysis has been applied to limited intrinsic frequency distributions, for instance, the Watanabe--Strogatz transform~\cite{Watanabe,moebius} for identical oscillators or the OA method for Lorentzian-type intrinsic frequency distributions~\cite{volcano,Hong11,Hong12,Ott08,Ott09,Martens09,vanHemmen,stefanovska,stefanovska_tw}. A general theory for the stability analysis is yet unknown. 

Note that the SC method does not determine the stability of the solution in general. The stability of each SC solution must be checked numerically. The empirical linear stability criterion proposed in \cite{stefanovska} has been used to identify the stability of the self-consistently obtained solutions in several cases with different intrinsic frequency distributions. However, this criterion is incomplete for metastable states with almost neutral stability for a uniform intrinsic frequency distribution~\cite{cpazo_pre}.

\section{The $c$-WP model}\label{sec:wp_model}
\begin{figure*}[!ht]
	\centering
	\includegraphics[width=0.7\textwidth]{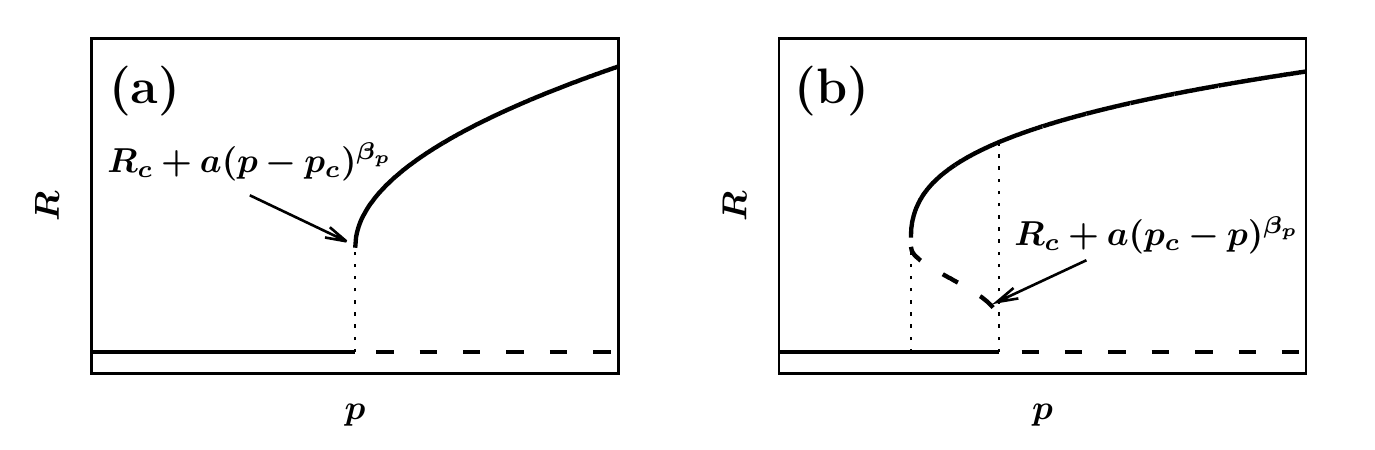}
	\caption{Schematic plots of the hybrid phase transition and the first-order transition with hysteresis: (a) a hybrid phase transition with $\beta_p=2/3$ occurs for $Q<1$, whereas (b) a first-order transition with hysteresis occurs for $Q > 1$, where an unstable branch with $\beta_p=2/3$ is hidden. Solid and dashed lines denote stable and unstable SC solutions, respectively, and the dotted vertical lines denote jumps at the transition points.}
	\label{fig:fig-hybrid-bifurcations}
\end{figure*}

Thus far, we have extended the SC framework to the $c$-KM. Here, we  
study the $c$-KM with a uniform intrinsic frequency distribution in the range $[-\gamma,\gamma]$ and two coupling constants, $K_1<0$ and $K_2>0$, at a ratio of $1-p$ to $p$. This model is the $c$-WP model. The distribution $g(K,\omega)$ is written as
\begin{equation}
g(K,\omega)=\frac{1}{2\gamma}\Theta(\gamma-|\omega|)\left[(1-p)\delta(K-K_1)+p\delta(K-K_2)\right].
\end{equation}
In fact, the KM with a uniform frequency distribution, which we called the Winfree-Paz\'{o} (WP) model, is known to exhibit a hybrid synchronization transition, which is a discontinuous phase transition~\cite{Winfree80} accompanying a singularity~\cite{Pazo05}. We also observe a hybrid phase transition as $p$ is varied in the $c$-WP model with a critical exponent $\beta_p=2/3$ at $p_c^+$, which is analogous to $\beta_K=2/3$ as $K$ is varied at $K_c^+$ in the WP model~\cite{Pazo05}. This transition occurs when $Q\equiv|K_1|/K_2<1$ [Fig.~\ref{fig:fig-hybrid-bifurcations}(a)]. Furthermore, when $Q>1$, there exists an unstable hybrid transition, as shown in Fig.~\ref{fig:fig-hybrid-bifurcations}(b).

When $\Omega=0$, i.e., for the IC and $\pi$ states, the imaginary part of Eq.~\eqref{eq:sce} vanishes because the integrand becomes odd in $\omega$. The transition from the IC to the $\pi$ state is characterized by solving the real part of the SC equation,
\begin{equation}
R = -(1-p)\int_{-|K_1|R}^{|K_1|R} d\omega g(\omega) \sqrt{1-\Big(\frac{\omega}{|K_1|R}\Big)^2}
+ p \int_{-K_2R}^{K_2R} d\omega g(\omega) \sqrt{1-\Big(\frac{\omega}{K_2R}\Big)^2},
\label{eq:sce-real}
\end{equation}
where $g(\omega)=\frac{1}{2\gamma}\theta(\gamma-|\omega|)$. In the following, analytic solutions of Eq.~\eqref{eq:sce-real} are obtained for different parameter values. Note that the integration is evaluated differently depending on the relative sizes of $|K_1|R$, $K_2R$, and $\gamma$. i) For $Q\equiv|K_1|/K_2<1$, $\gamma$ can fall in one of the three ranges $(0,|K_1|R\,]$, $(|K_1|R,K_2R\,],$, and $(K_2R,\infty)$. ii) For $Q>1$, $\gamma$ can fall in one of the three ranges $(0,K_2R\,]$, $(K_2R,|K_1|R\,],$, and $(\,|K_1|R,\infty)$.

\subsection{The SC solution for IC and $\pi$}
\subsubsection{Supercritical hybrid phase transition ($Q<1$)\\} 
The integration of Eq.~\eqref{eq:sce-real} when $Q<1$ yields
\begin{equation}
R= \begin{cases}
-\frac{1-p}{2\gamma} \frac{\pi}{2}|K_1| R +\frac{p}{2\gamma}\frac{\pi}{2}K_2 R & {\rm for~~} R <\frac{\gamma}{|K_1|}< \frac{\gamma}{K_2}\\[10pt]  
-(1-p)\Big(\frac{|K_1|R}{2\gamma}\arcsin\frac{\gamma}{|K_1|R} +\frac{1}{2}\sqrt{1-\big(\frac{\omega}{|K_1|r}\big)^2}\Big)+\frac{p}{2\gamma}\frac{\pi}{2}K_2 R & {\rm for~~} \frac{\gamma}{|K_1|}< R <\frac{\gamma}{K_2} \\[10pt]
-(1-p)\Big(\frac{|K_1|R }{2\gamma}\arcsin\frac{\gamma}{|K_1| R } +\frac{1}{2}\sqrt{1-\big(\frac{\omega}{|K_1|R}\big)^2}\Big) \\
\hspace{9em}+ p\Big(\frac{K_2 R}{2\gamma}\arcsin\frac{\gamma}{K_2R} +\frac{1}{2}\sqrt{1-\big(\frac{\omega}{K_2R}\big)^2}\Big) & {\rm for~~} \frac{\gamma}{|K_1|}< \frac{\gamma}{K_2} < R.
\end{cases}
\end{equation}
Note that the IC ($R=0$) state is a trivial solution of the above SC equation. The remaining nontrivial solutions correspond to $\pi$ states. After inverting the above equations and solving for $p$, we obtain the inverse function of the order parameter curve $p(R)$ as follows:
\begin{equation}
p = \begin{cases}
\frac{|K_1|+4\gamma/\pi}{|K_1|+K_2}  & {\rm for~~} 0< R < \frac{\gamma}{K_2}, \\[13pt]
\frac{(\pi |K_1|+4 \gamma) R }{\pi |K_1| R +2 \gamma  \sqrt{1-(\frac{\gamma}{K_2 R })^2}+2 K_2 R \arcsin(\frac{\gamma }{K_2 R})} &{\rm for~~}  \frac{\gamma}{K_2}< R <\frac{\gamma}{|K_1|}, \\[15pt]
\frac{\gamma  \sqrt{1-(\gamma/|K_1|R)^2}+ |K_1| R \arcsin\left(\frac{\gamma }{|K_1| R}\right)+2 \gamma  R}{\gamma  \sqrt{1-(\frac{\gamma}{|K_1|R})^2}+|K_1| R \arcsin\left(\frac{\gamma }{|K_1| r}\right)+\gamma \sqrt{1-(\frac{\gamma}{K_2R})^2}+K_2 R \arcsin\left(\frac{\gamma }{K_2 R}\right)} &{\rm for~~}  \frac{\gamma}{K_2}< \frac{\gamma}{|K_1|} < R.
\end{cases} \label{eq:supercriticalQ-ISCE}
\end{equation}
Numerical solutions of the order parameter curve $R(p)$ from the SC equations were presented in \cite{cpazo_pre}. Moreover, the stability of the SC solutions was determined by linear stability analysis~\cite{stefanovska}, which will be discussed later.

The order parameter curve $R(p)$ shows a discontinuous jump of size $R_c$ at the critical point $p_c$.
\begin{eqnarray}
p_c &=\frac{|K_1|+4\gamma/\pi}{|K_1|+K_2} = \frac{Q+\frac{4\gamma}{\pi K_2}}{Q+1} , \nonumber \\
R_c &=\gamma/K_2.
\end{eqnarray}
Expanding the intermediate branch $\frac{\gamma}{|K_1|} < R < \frac{\gamma}{K_2}$ in powers of $\epsilon\equiv (R-R_c)/{R_c}$ after the jump gives
\begin{eqnarray}
p&=\frac{(\pi |K_1|+4 \gamma)R}{\pi |K_1| R +2 \gamma  \sqrt{1-\big(\frac{\gamma}{K_2R}\big)^2} +2 K_2 R \arcsin \big(\frac{\gamma }{K_2 R}\big)} \nonumber \\[10pt]
&\approx p_c + \frac{8\sqrt{2}}{3\pi}\frac{p_c}{(Q+1)} \epsilon^{3/2} + O(\epsilon^{5/2}).
\end{eqnarray}
Therefore, the transition from the IC to the $\pi$ state is hybrid, as $(R-R_c)\sim(p-p_c)^{\beta_p}$, with $\beta_p=2/3$ for $Q<1$. Note that at $p_c$, the competing system has a mean coupling strength
\begin{eqnarray}
\langle K \rangle_c = (1-p_c)K_1 + p_c K_2 = \frac{4\gamma}{\pi},
\end{eqnarray}
which is equivalent to the critical coupling strength $K_c=2/\pi g(0) = 4\gamma/\pi$ for the IC-to-C transition in the WP model~\cite{Winfree80,Pazo05}. Note also the appearance of the critical exponent value for the WP model, $\beta_K=2/3$. This is rather natural because the WP model corresponds to a particular case of the $c$-WP model with $p=1$. In addition, successive transitions from the IC to the $\pi$ and then to the TW state are observed in some parameter regimes. Further details are provided in \cite{cpazo_pre}.

\subsubsection{Subcritical hybrid phase transition ($Q>1$)\\}
The integration of Eq.~\eqref{eq:sce-real} when $Q>1$ yields
\begin{equation}
R= \begin{cases}
-\frac{1-p}{2\gamma} \frac{\pi}{2}|K_1| R +\frac{p}{2\gamma}\frac{\pi}{2}K_2 R & {\rm for~~} R <\frac{\gamma}{|K_1|}< \frac{\gamma}{K_2}\\[10pt]  
-(1-p)\Big(\frac{|K_1|R}{2\gamma}\arcsin\frac{\gamma}{|K_1|R} +\frac{1}{2}\sqrt{1-\big(\frac{\omega}{|K_1|r}\big)^2}\Big)+\frac{p}{2\gamma}\frac{\pi}{2}K_2 R & {\rm for~~} \frac{\gamma}{|K_1|}< R <\frac{\gamma}{K_2} \\[10pt]
-(1-p)\Big(\frac{|K_1|R }{2\gamma}\arcsin\frac{\gamma}{|K_1| R } +\frac{1}{2}\sqrt{1-\big(\frac{\omega}{|K_1|R}\big)^2}\Big) \\
\hspace{8em}+ p\Big(\frac{K_2 R}{2\gamma}\arcsin\frac{\gamma}{K_2R} +\frac{1}{2}\sqrt{1-\big(\frac{\omega}{K_2R}\big)^2}\Big) & {\rm for~~} \frac{\gamma}{|K_1|}< \frac{\gamma}{K_2} < R.
\end{cases}
\end{equation}

\begin{figure}[!hbt]
	\centering
	\includegraphics[width=\linewidth]{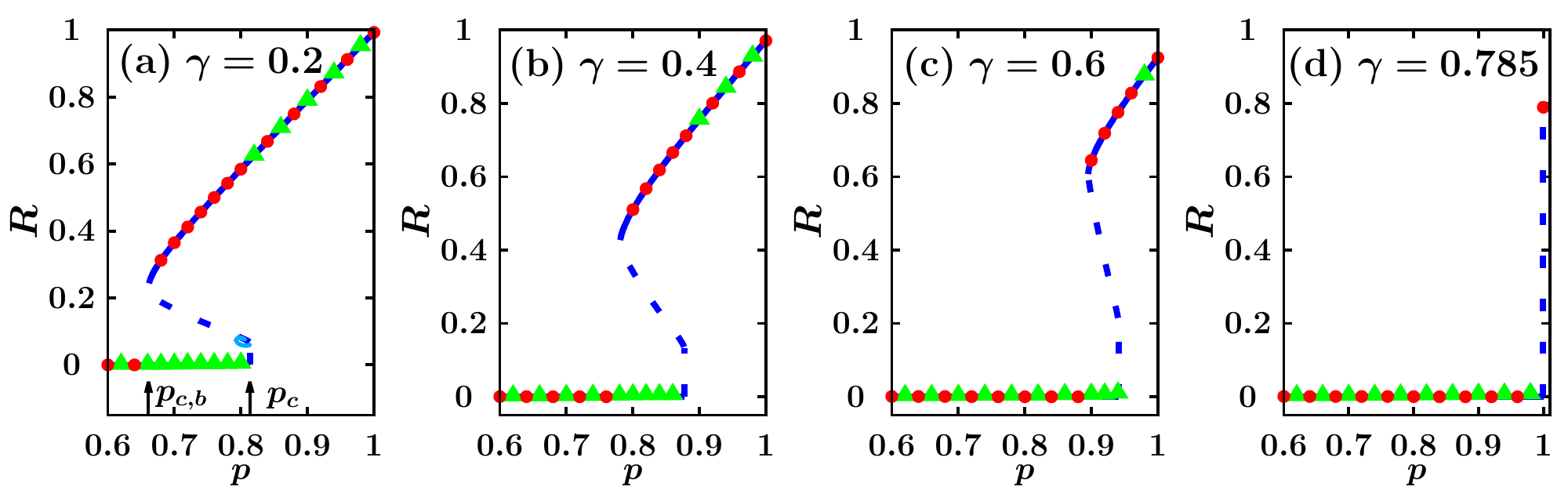}
	\caption{Plot of the order parameter versus $p$ for $Q=3$ and various $\gamma$ values: (a) $\gamma=0.2$, (b) $\gamma=0.4$, (c) $\gamma=0.6$, and (d) $\gamma_h=\pi K_2/4\approx0.785$. A first-order transition and hysteresis occur between $p_c$ and $p_{c,b}$. With increasing $\gamma$, $p_{c,b}$ and $p_c$ increase at different rates, and the hysteresis range is reduced. The unstable solution contains a discontinuous jump and a hidden exponent, $\beta_p=2/3$. Note also that in (d), at $\gamma=\gamma_{h}=\pi K_2/4\approx 0.785$, $p_c^b=p_c=1$, and $R_c=\gamma/K_2$. At $p=1$, the $c$-WP model reduces to the WP model~\cite{Winfree80,Pazo05}. For $\gamma > \gamma_{h}$, the system is IC because the value of $K_2$ is subcritical; i.e., $K_2< K_c \equiv 4\gamma/\pi$, where $K_c$ is the critical coupling strength of the WP model.}
	\label{fig:r_p_large}
\end{figure}

Similarly, we obtain 
\begin{equation}
p = \begin{cases} 
\frac{|K_1|+4\gamma/\pi}{|K_1|+K_2}  & {\rm for~~} 0 < R <\frac{\gamma}{|K_1|}\\[11pt]
\frac{2R\gamma + \gamma\sqrt{1-(\frac{\gamma}{|K_1|R})^2}  + |K_1|R\arcsin\left(\frac{\gamma}{|K_1|R}\right)}{\frac{\pi K_2 R}{2} +\gamma\sqrt{1-(\frac{\gamma}{|K_1|R})^2 }  + |K_1|R\arcsin\left(\frac{\gamma}{|K_1|R}\right)} & {\rm for~~} \frac{\gamma}{|K_1|}<R<\frac{\gamma}{K_2} \\[11pt]
\frac{\gamma \sqrt{1-(\frac{\gamma}{|K_1|R})^2}  + |K_1|R\arcsin\left(\frac{\gamma}{|K_1|R}\right)+2 \gamma R}{\gamma \sqrt{1-(\frac{\gamma}{|K_1|R})^2}+|K_1|R\arcsin\left(\frac{\gamma }{|K_1| R}\right)+\gamma  \sqrt{1-(\frac{\gamma}{K_2R})^2}+K_2 R \arcsin\left(\frac{\gamma }{K_2 R}\right)} & {\rm for~~}  \frac{\gamma}{|K_1|}< \frac{\gamma}{K_2}< R. 
\end{cases} \label{eq:subcriticalQ-ISCE}
\end{equation}
As $R\rightarrow R_c={\gamma}/{|K_1|}$, for $\epsilon = (R_{c}-R)/R_{c}$:
\begin{eqnarray}
p&=\frac{2R\gamma + \gamma\sqrt{1-(\frac{\gamma}{|K_1|R})^2}  + |K_1|R\arcsin\left(\frac{\gamma}{|K_1|R}\right)}{\frac{\pi K_2 R}{2} +\gamma\sqrt{1-(\frac{\gamma}{|K_1|R})^2 }  + |K_1|R\arcsin\left(\frac{\gamma}{|K_1|R}\right)}\nonumber \\[10pt]
&\approx\frac{2\gamma R_c (1+\epsilon)
+\gamma\left[\sqrt{2\epsilon}-\frac{3\epsilon^{3/2}}{2\sqrt{2}}+\cdots\right]
+\gamma(1+\epsilon)\left[\frac{\pi}{2}-\sqrt{2\epsilon}+\frac{5 \epsilon ^{3/2}}{6\sqrt{2}}-\cdots\right]}{\frac{\pi K_2}{2} R_c(1+\epsilon)
+\gamma\left[\sqrt{2\epsilon}-\frac{3 \epsilon ^{3/2}}{2 \sqrt{2}}+\cdots\right]
+\gamma(1+\epsilon)\left[\frac{\pi}{2}-\sqrt{2\epsilon}+\frac{5 \epsilon^{3/2}}{6 \sqrt{2}}-\cdots\right]} \nonumber \\[10pt]
&\approx \frac{\pi |K_1|+4 \gamma}{\pi |K_1|+\pi K_2} - \frac{8\sqrt{2}(\pi K_2 -4 \gamma)\gamma R_c}{3\pi^2(K_2 R_c + \gamma)^2}\epsilon^{3/2}+O(\epsilon^{5/2}). 
\end{eqnarray}
As the model approaches the IC-to-$\pi$ transition for $Q>1$, there is an unstable solution that is separated from the IC phase, as shown in Fig.~\ref{fig:r_p_large}. Here, the unstable solution follows $(R-R_c)\sim(p_c-p)^{2/3}$, which gives the same exponent, $\beta_p=2/3$; however, $p_c$ is approached from the opposite direction $p\to p_c^-$. In fact, this hybrid transition for $Q > 1$ is hidden in the simulations, because it corresponds to an unstable branch. The discontinuous transition for $Q>1$ shows a hysteresis curve that starts at $p_{c,f}=p_c$ in the forward direction and at $p_{c,b}$ in the backward direction, where $p_{c,f}$ is given as 
\begin{equation}
p_{c,f} = \frac{|K_1|+4\gamma/\pi}{|K_1|+K_2}  = \frac{Q+\frac{4\gamma}{\pi K_2}}{Q+1} =p_c,
\end{equation}
and $p_{c,b}$ and $R_{c,b}$ are determined numerically using the equation
\begin{equation}
\frac{dp}{dR}
= \frac{d}{dR}\left( \frac{\gamma \sqrt{1-(\frac{\gamma}{|K_1|R})^2}  + |K_1|R\arcsin\left(\frac{\gamma}{|K_1|R}\right)+2 \gamma R}{\gamma \sqrt{1-(\frac{\gamma}{|K_1|R})^2}+|K_1|R\arcsin\left(\frac{\gamma }{|K_1| R}\right)+\gamma  \sqrt{1-(\frac{\gamma}{K_2R})^2}+K_2 R \arcsin\left(\frac{\gamma }{K_2 R}\right)} \right)=0.
\end{equation}

\subsection{Phase diagram}
\begin{figure}[!ht]
	\centering
	\includegraphics[width=\linewidth]{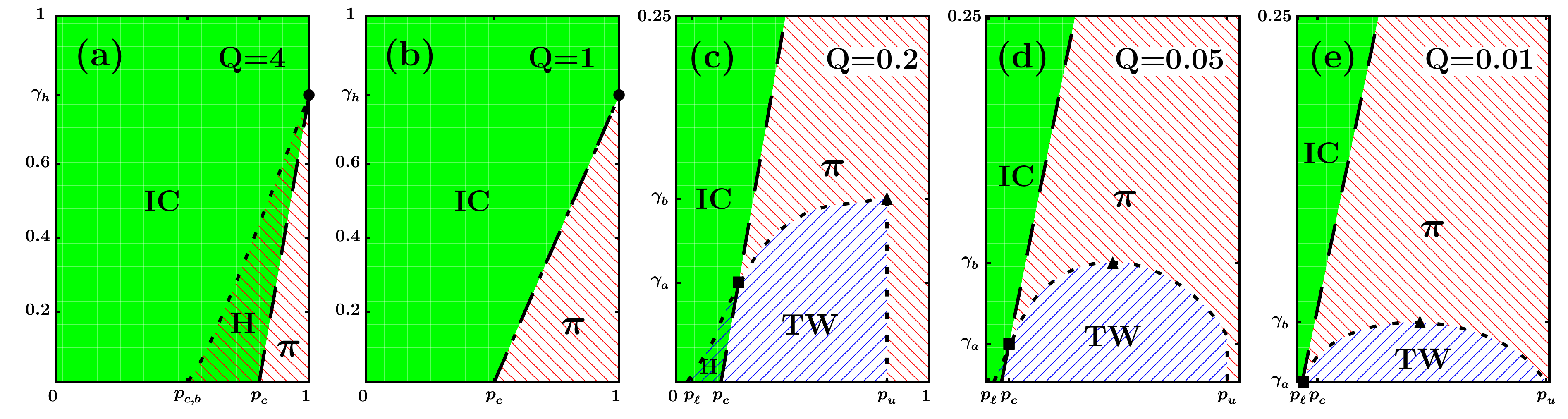}
	\caption{Phase diagram of synchronization transitions in the $(p,\gamma)$ plane for various $Q$. Here, $p$ is the fraction of oscillators with positive coupling $K_2$, and $\gamma$ is the half-width of the uniform distribution $g(\omega)$. (a) When $Q>1$, the phase diagrams contain IC and $\pi$ phases and the hysteresis zone {\bf H} of the two phases. Both types of dashed lines represent discontinuous transitions; however, the forward transition from the IC to the $\pi$ phase is hybrid. (b) At $Q=1$, the hysteresis vanishes. The $p=1$ line corresponds to the phase diagram of the WP model, and the symbol $\bullet$ at $\gamma_h=\pi K_2/4\approx 0.78$ denotes the hybrid critical synchronization transition point for $K_2=1$~\cite{Winfree80,Pazo05}. (c)$-$(e) When $Q < 1$, some part of the $\pi$ state is replaced by the TW phase. The hysteresis region reappears between the IC and TW phases. The symbols $\blacktriangle$ and $\blacksquare$ represent critical points across which different types of phases or phase transitions emerge.}
	\label{fig:phaseDiagram}
\end{figure}

Phase diagrams of the synchronization transitions in the $(p,\gamma)$ plane for various $Q$ values are presented in Fig.~\ref{fig:phaseDiagram}. In Figs.~\ref{fig:phaseDiagram}(a) and (b), we consider the case $Q\geq 1$. The  phase diagrams contain IC and $\pi$ phases and the hysteresis zone {\bf H} of the two phases. Both types of dashed lines represent discontinuous transitions; however, the forward transition from the IC to the $\pi$ phase is hybrid. At $Q=1$ in Fig.~\ref{fig:phaseDiagram}(b), the hysteresis vanishes. The $p=1$ line corresponds to the phase diagram of the WP model, and a hybrid synchronization transition occurs at $\gamma_h\approx 0.78$ ($\bullet$) in the WP model~\cite{Winfree80,Pazo05}. In Figs.~\ref{fig:phaseDiagram}(c)$-$(e), we consider the case $Q < 1$. Part of the $\pi$ state is occupied by the TW phase. The hysteresis region reappears between the IC and TW phases. As the ratio $Q$ is decreased, the interval $[p_{\ell},p_u]$ of the TW state becomes broader. 

\subsection{The empirical stability of the SC solutions}
To characterize the stability of the SC solutions, Iatsenko et al.~\cite{stefanovska} assumed a linear relaxation dynamics of the complex order parameter for small perturbations near each SC solution. The (empirical) stability matrix $\hat{S}$ is reproduced as follows:
\begin{eqnarray}
\begin{pmatrix} \dot{\delta R} \\ \dot{\delta \psi} \end{pmatrix}
&= A \begin{pmatrix}(\partial_R F_R)-1 & R^2 \partial_\Omega F_R \\ R^{-1}\partial_R F_\Omega & R\partial_\Omega F_\Omega\end{pmatrix} \begin{pmatrix} \delta R \\ \delta \psi \end{pmatrix} \nonumber \\
&\equiv A \hat{S}\begin{pmatrix} \delta R \\ \delta \psi \end{pmatrix},
\end{eqnarray}
where $F_R$ and $F_\Omega$ are the real and imaginary parts of the SC complex order parameter, respectively. They are written explicitly as 
\begin{eqnarray}
F_R(R,\Omega) &\equiv \int_{\rm locked} dKd\omega g(\omega,K) 
\sqrt{1-\left(\omega/KR\right)^2}, \nonumber \\
F_\Omega(R,\Omega) &\equiv \int_{\rm drifting} dKd\omega g(\omega,K) 
\sqrt{\left(\omega/KR\right)^2-1}.
\label{eq:stability}
\end{eqnarray}
In the stability matrix $\hat{S}$, the relative strength of the proportional coefficients of $\delta R$ and $\delta \Omega$ has been determined empirically~\cite{stefanovska}. 

The empirical stability of the SC solutions of the $c$-WP model is represented by the blue solid (stable) and dashed (unstable) curves in Fig.~\ref{fig:r_p_large} and Fig. 2 of \cite{cpazo_pre}. Our numerical result suggests that for neutrally stable solutions, the linear stability criterion of \cite{stefanovska} may be partially fulfilled. In our previous study~\cite{cpazo_pre}, we found neutrally (or weakly) stable $\pi$ solutions, which are stable in the $R$ direction but (nearly) neutral in the $\Omega$ direction. Thus, the order parameter trajectories exhibited some drifting motion near these marginally stable solutions. These solutions characterize metastable states, which capture the system for a long time; however, the system eventually escapes toward the final stable TW solutions.

\section{$c$-WP model on ER networks}\label{sec:er}
\subsection{The SC equation for ER networks}\label{sec:er_anneal}
Now we generalize the $c$-KM to random networks. The Kuramoto equation is written as 
\begin{eqnarray}
\dot\theta_i &= \omega_i + \frac{K_i}{\langle q\rangle} \sum_{j=1}^{N}a_{ij} \sin(\theta_j-\theta_i),
\label{eq:ER-cKM}
\end{eqnarray}
where the normalization $\langle q\rangle$ is equivalent to the expected number of terms in the summation, and $a_{ij}$ is the adjacency matrix. On the ER network, $a_{ij}=1$ if two nodes $i$ and $j$ are connected; otherwise, $a_{ij}=0$. For all-to-all connections, this equation reduces to Eq.~\eqref{eq:km}.
Now we consider an annealed ER network, in which
\begin{eqnarray}
a_{ij} = \frac{q_i q_j}{N\langle q\rangle}.
\label{eq:annealed}
\end{eqnarray}
Here, $q_i$ and $q_j$ are the degrees of nodes $i$ and $j$, respectively. Each degree follows a Poisson distribution,
\begin{eqnarray}
P_d(q) = \frac{\langle q\rangle^q e^{-\langle q\rangle}}{q!},
\end{eqnarray}
where again $\langle q\rangle$ is the mean degree. When the order parameter is defined as the degree-weighted phase coherence,
\begin{eqnarray}
H e^{{\rm i}\phi} = \frac{\sum_j q_j e^{{\rm i}\theta_j}}{\sum_j q_j} = \frac{1}{N\langle q\rangle}\sum_j q_j e^{{\rm i}\theta_j},
\end{eqnarray}
the model is written as  
\begin{eqnarray}
\dot\theta_i &= \omega_i + \frac{K_i q_i H}{\langle q \rangle} \sin(\phi-\theta_i),
\label{eq:ER-cKM-MF}
\end{eqnarray}
where we take $\phi=0$ for simplicity. In the stationary limit, the oscillators satisfying $|\omega_i-\Omega| \leq |K_i| q_i H/\langle q\rangle$ are phase-locked, where $\Omega$ denotes the angular velocity of the complex order parameter $He^{i\phi}$. Each locked oscillator contributes a phasor of 
\begin{eqnarray}
e^{{\rm i}\theta_j} = \sqrt{1-\left(\frac{\langle q\rangle(\omega_j-\Omega)}{K_jq_jH}\right)^2}+{\rm i}\frac{\langle q\rangle(\omega-\Omega)}{K_jq_jH},
\end{eqnarray}
and each drifting oscillator rotating with a period
\begin{eqnarray}
T_j = \int_0^{2\pi}\frac{d\theta_j}{|\dot\theta_j|} = \frac{2\pi}{\sqrt{(\omega_i-\Omega)^2-(K_jq_jH/\langle q \rangle)^2}}
\end{eqnarray}
makes a time-averaged contribution of
\begin{eqnarray}
\overline{e^{i\theta_j}}&= \frac{1}{T_j}\int_0^{2\pi} \frac{d\theta_j}{\dot\theta_j} e^{i\theta_j} \nonumber \\
&= \frac{i\langle q \rangle\textrm{sgn}(\omega_j-\Omega)}{K_jq_jH}\left[|\omega_j-\Omega|-\sqrt{(\omega_j-\Omega)^2-(K_jq_jH/\langle q \rangle)^2}\right].
\end{eqnarray}
Thus, the SC equation of the $c$-KM on an ER network is written as 
\begin{eqnarray}
H &= \int d\omega dK g(\omega,K) \frac{\sum_q q P_d(q) \overline{e^{{\rm i}\theta_j}}}{\sum_q q P_d(q)} \nonumber \\
&= \int d\omega dK g(\omega,K) \frac{1}{\langle q\rangle} \bigg[\sum_{q<|X|} q P_d(q) \overline{e^{{\rm i}\theta}} + \sum_{q>|X|} qP_d(q) e^{{\rm i}\theta}\bigg] \nonumber \\
&= \int d\omega dK g(\omega,K) \frac{1}{\langle q\rangle} \bigg[\sum_{q>|X|} P_d(q) \sqrt{q^2-X^2} \nonumber \\
&~~~~~~~~~~~~~~~~~~~~~~+ {\rm i} X - {\rm i}~\textrm{sgn}(X) \sum_{q<|X|} P_d(q)\sqrt{X^2-q^2}\bigg],
\label{eq:ER-cKM-SCE}
\end{eqnarray}
where we used the shorthand notation $X=\langle q \rangle(\omega-\Omega)/(KH)$.
\begin{figure}[!t]
	\centering
	\includegraphics[width=0.6\linewidth]{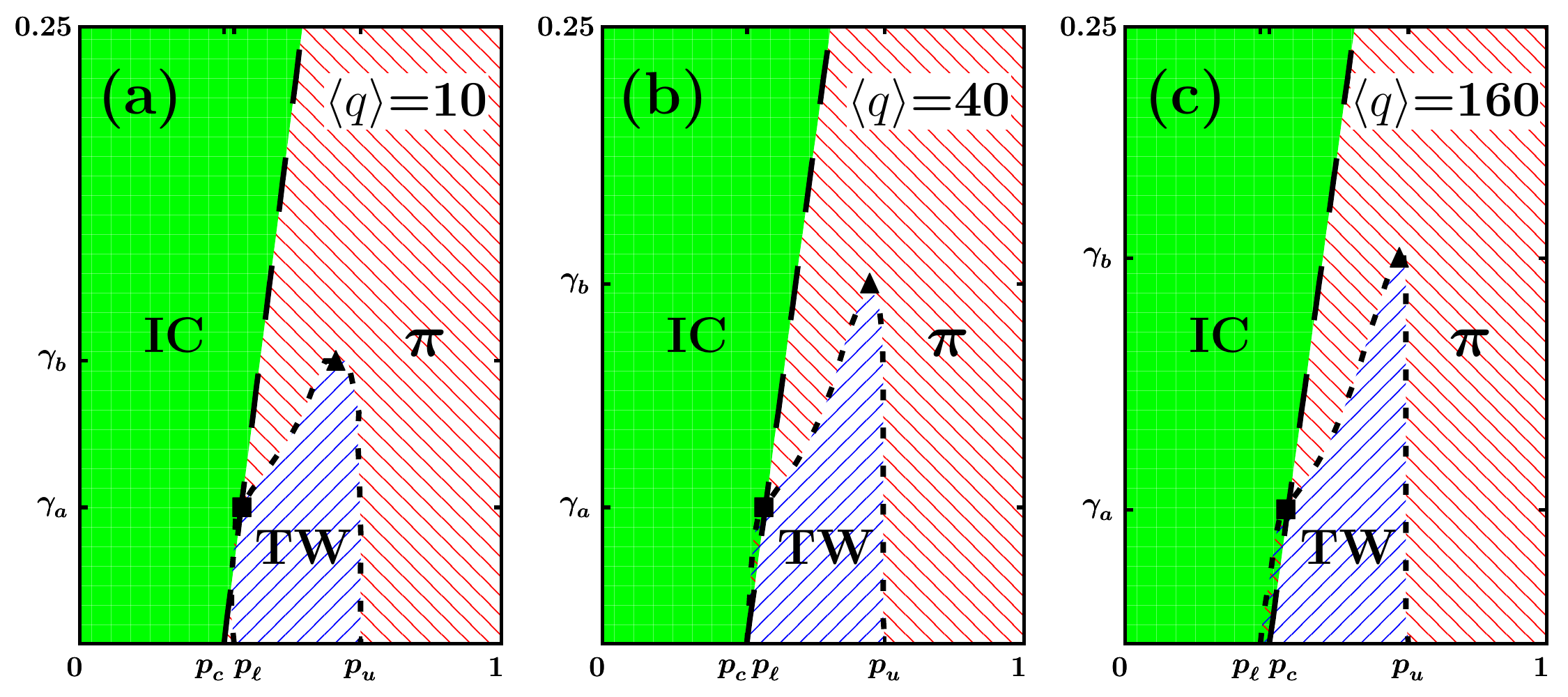}
	\caption{Phase diagram of the $c$-WP model on the annealed ER networks $a_{ij}=q_iq_j/N\langle q\rangle$ drawn on the $(p,\gamma)$ plane for $\langle q\rangle=10,40$, and $160$. Here, $p$ is the fraction of oscillators with a positive coupling constant $K_2$, and $\gamma$ is the half-width of the uniform distribution $g(\omega)$. $Q=|K_1|/K_2$ is fixed at 0.5. The TW regime expands as the mean degree $\langle q\rangle$ is increased, and finally Fig.~\ref{fig:phaseDiagram}(c) is obtained as $\langle q\rangle\rightarrow N$.}
	\label{fig:fig-ER-PD}
\end{figure}

Eq.~\eqref{eq:ER-cKM-SCE} can be numerically solved. The corresponding mean-field phase diagram is shown in Fig.~\ref{fig:fig-ER-PD}. The TW regime expands as the network mean degree $\langle q\rangle$ is increased. We confirm that as $\langle q\rangle\rightarrow N$, the solution is reduced to the solution on the all-to-all network shown in Fig.~\ref{fig:phaseDiagram}(c). The stability of these mean-field solutions was tested by numerical simulations on the annealed ER network governed by Eq.~\eqref{eq:ER-cKM-MF}. The black symbols in Fig.~\ref{fig:fig-ER} represent the stationary states of the simulation, which were reached after waiting a sufficiently long time. The solid (dotted) blue curves in Fig.~\ref{fig:fig-ER} denote stable (unstable) solutions of the SC equations under the linear empirical stability criterion, which was calculated by a method similar to one in the literature \cite{cpazo_pre,stefanovska}. We observe that the TW and $\pi$ phases also appear in the annealed ER network. In addition, the empirical stability condition applies in this mean-field solution in most of the regime; however as before, metastability does arise in the $\pi$ states (solid blue), as they are covered by the simulation data points only for an intermediate range of time. Therefore, the mean-field theory is as successful for annealed ER networks as the SC theory is for all-to-all networks in terms of investigating the synchronization phase transition in the $c$-WP model.

\subsection{Synchronization transitions of $c$-WP model on ER networks}\label{sec:er_simulation}
\begin{figure*}[!ht]
	\centering
	\includegraphics[width=\linewidth]{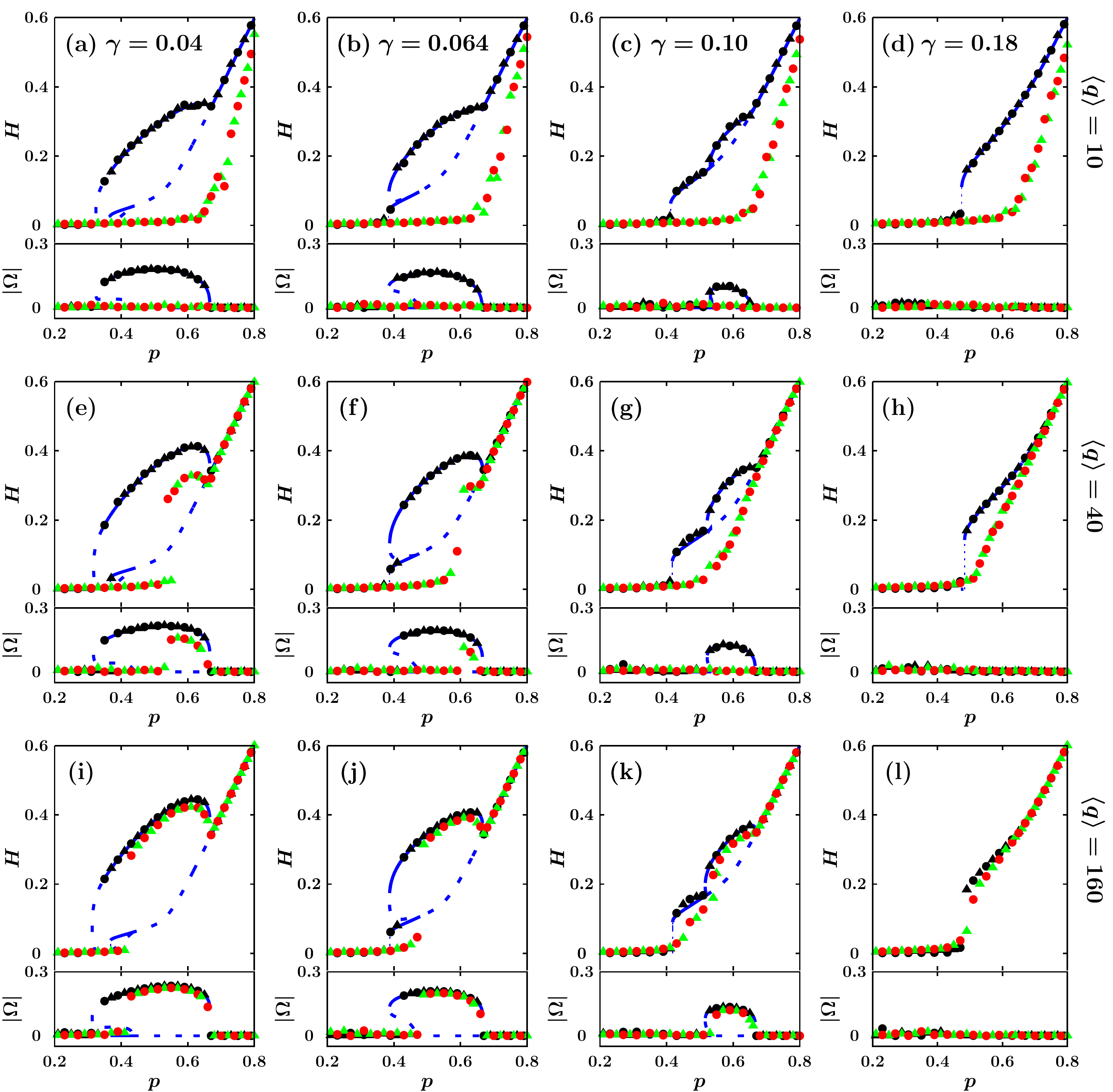}
	\caption{Order parameter curves for the $c$-WP model on ER networks. $N=10^4$, $Q=0.5$, $\gamma=0.04,0.064,0.10$, and $0.18$ (from left to right), and $\langle q\rangle=10,40$, and $160$ (from top to bottom). The blue curves and black data points represent the self-consistency solutions and the simulation results of the annealed ER network $a_{ij}=q_i q_j/N\langle q\rangle$. Red ($\bullet$) and green ($\blacktriangle$) symbols denote simulation results for the ER network starting from IC and C initial states, respectively. Each red/green data point corresponds to a single random network realization. Synchronization from the IC to the TW or $\pi$ state also occurs in the ER network. Here, the phase transition depends on not only the frequency half-width $\gamma$ but also on the mean degree $\langle q\rangle$. Interestingly, the TW phase is not observed for the ER network with a relatively low mean degree $\langle q\rangle\approx 10$. For a dense ER network with a large mean degree ($\langle q\rangle\geq 160$), the result is relatively in good agreement with the mean-field calculaions.}
	\label{fig:fig-ER}
\end{figure*}

Next, we compare the results of numerical simulations of $c$-KM on quenched ER random networks, which use the conventional adjacency matrix~\eqref{eq:ER-cKM}. In Fig.~\ref{fig:fig-ER}, the simulation data points of networks starting from a C/IC state are denoted by green ($\blacktriangle$)/red ($\bullet$) symbols. A comparison with the mean-field results on the annealed ER networks reveals similarities and differences. The TW and $\pi$ phases also appear in the $c$-KM model on ER networks, but only for a sufficiently large mean degree $\langle q\rangle$. Moreover, the order of transition may depend on $\langle q\rangle$. In the ER networks with low mean degree [Figs.~\ref{fig:fig-ER}(a)$-$(d)], the TW is not observed. The synchronization transition to the $\pi$ phase requires a fraction $p$ of $K_2$ oscillators that is larger than $p_c\approx 0.66$, and the transition is continuous; however, the $p_c$ value in the mean-field result is smaller, and the transition can be discontinuous. For an intermediate mean degree [Figs.~\ref{fig:fig-ER}(e)$-$(h)], we observe the emergence of a TW phase, as in the mean-field prediction; however, the order of transition depends on the frequency half-width $\gamma$. 
In Fig.~\ref{fig:fig-ER}(e), we unexpectedly find that $p$ exhibits hysteresis behavior in the range $[0.54,0.55]$, which did not occur in the mean-field calculations and simulations. Note that in Fig.~\ref{fig:fig-ER}(g), the TW regime appears, although the angular velocity is very small. For higher mean degree [Figs.~\ref{fig:fig-ER}(i)$-$(l)], the hybrid jump transition at $p_c$ is gradually restored.

The mean-field theory of the $c$-WP model works for the annealed ER networks~\eqref{eq:annealed}. Note, however, that the annealed ER networks are weighted and fully connected. By contrast, the synchronization phase transition of the $c$-WP model on the quenched ER networks differs significantly, especially when the mean degree $\langle q\rangle$ is small. In the $c$-WP model, the annealed network approximation fails for the quenched ER network, in contrast to the case of ordinary KMs on complex networks~\cite{kuramoto-complex}. The main reason for this failure is the heterogeneity of the local environment, which is caused by the network disorder and the oscillators with a negative coupling constant $K_1$. In particular, for a small mean degree $\langle q\rangle$, the number of connections to the neighboring $K_2$ nodes is even smaller. Hence, a local order parameter is subjected to large finite-$\langle q\rangle$ fluctuations, and the sign of $K$ is dominant in the competing synchronization.
\begin{figure}[!t]
	\centering
	\includegraphics[width=\linewidth]{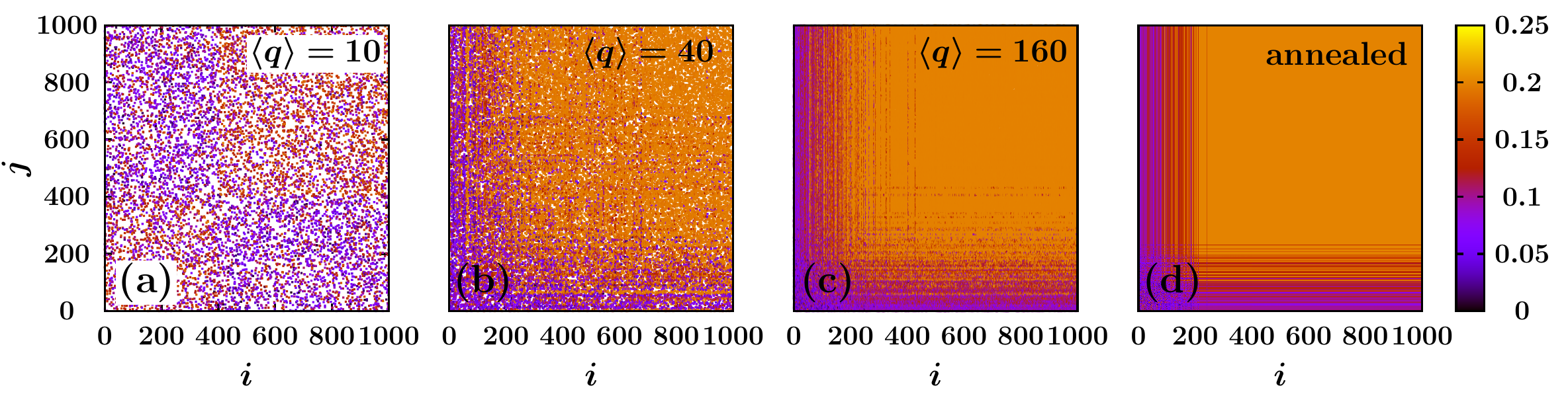}
	\caption{Link coherence $R_{ij}\equiv a_{ij}\left|\frac{1}{T}\int_0^T e^{i[\theta_i(t)-\theta_j(t)]}dt\right|$ between nodes $i$ and $j$, where $a_{ij}=1$ when $i,j$ are connected and $=0$ when they are disconnected. Simulations were performed at $N_1=400$, $N_2=600$, $\gamma=0.064$, for a single network realization each of (a) $\langle q\rangle=10$, (b) $\langle q\rangle=40$, and (c) $\langle q\rangle=160$, which correspond to $p=0.6$ in Figs.~\ref{fig:fig-ER}(b), (f), and (j), respectively. In (d), an annealed network with $\langle q\rangle=160$ was used. Results were time-averaged over $T=10^3$. Oscillator indices $i$ and $j$ are presented in increasing order of coupling constant and natural frequency, i.e., $\omega\approx-\gamma$ to $\omega\approx\gamma$ at $i=0$ to $399$ (for $K_i=K_1=-0.5$) and at $i=400$ to $999$ (for $K_i=K_2=1$).}
	\label{fig:fig-correlation}
\end{figure}
To investigate this effect, we first measured the link coherence $R_{ij}\equiv a_{ij}\left|\frac{1}{T}\int_0^T e^{i[\theta_i(t)-\theta_j(t)]}dt\right|$~\cite{moreno_2007}, as shown in Fig.~\ref{fig:fig-correlation}. For a sparse connection, $\langle q\rangle=10$ [Fig.~\ref{fig:fig-correlation}(a)], the link coherence matrix can be blocked; for the intraspecies interaction between $K_1$ and $K_1$ oscillators and $K_2$ and $K_2$ oscillators, coherences are high, and for the interspecies interaction between $K_1$ and $K_2$ oscillators, coherence is low. In addition, the values within the block are random and highly heterogeneous. The link coherence of the $K_1$ and $K_2$ oscillators is distributed within a broad range of values. Such large fluctuations can cause the mean-field theory to break down. As the mean number of neighbors $\langle q\rangle$ increases, the border between $K$-species gradually fades, and the link coherence among $K_2$ oscillators eventually becomes homogeneous. However, noisy features remain even when $\langle q\rangle=160$. By contrast, the link coherence of the fully connected, annealed ER network exhibits better organization, as shown in Fig.~\ref{fig:fig-correlation}(d). Note that the link coherence has a homogeneous value over an extended area, including all of the $K_2$ population and some of the $K_1$ population on the high-frequency sides. The corresponding oscillators have been entrained into a TW cluster. In addition, the low-coherence striped patterns correspond to some $K_1$ oscillators that are very weakly entrained with the TW cluster. For those oscillators, the density of connections $q_i$ is small and/or the natural frequency values are distant from the TW velocity. They remain detrained from the TW cluster.

\begin{figure}[!ht]
	\centering
	\includegraphics[width=\linewidth]{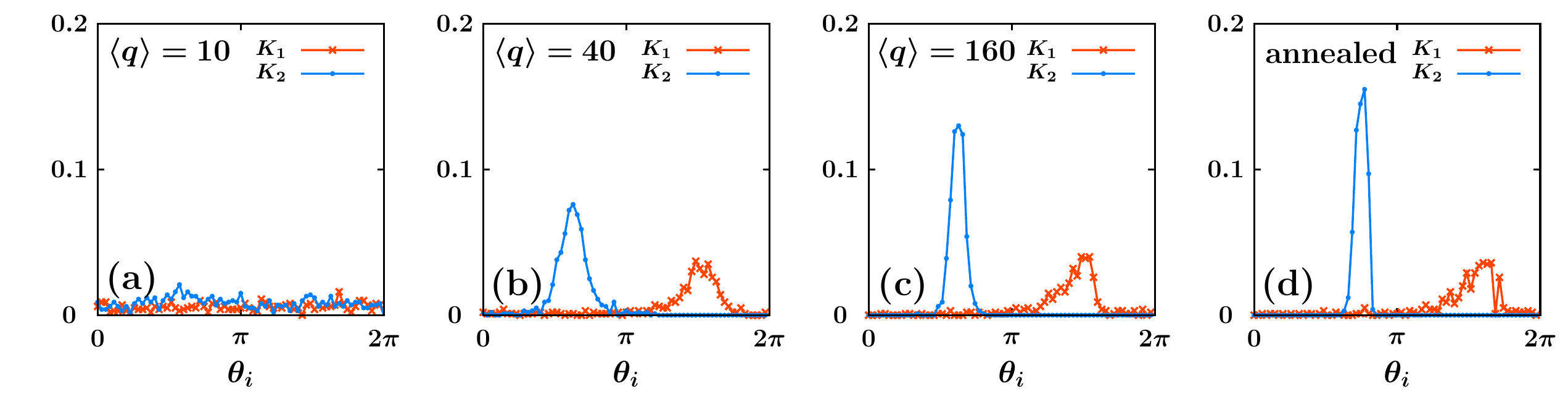}
	\caption{ Histogram of the oscillator phases. Simulation conditions are the same as in Fig.~\ref{fig:fig-correlation}. Bin size is $\Delta\theta=2\pi/70$.}
	\label{fig:fig-phase-scatter}
\end{figure}
\begin{figure}[!ht]
	\centering
	\includegraphics[width=\linewidth]{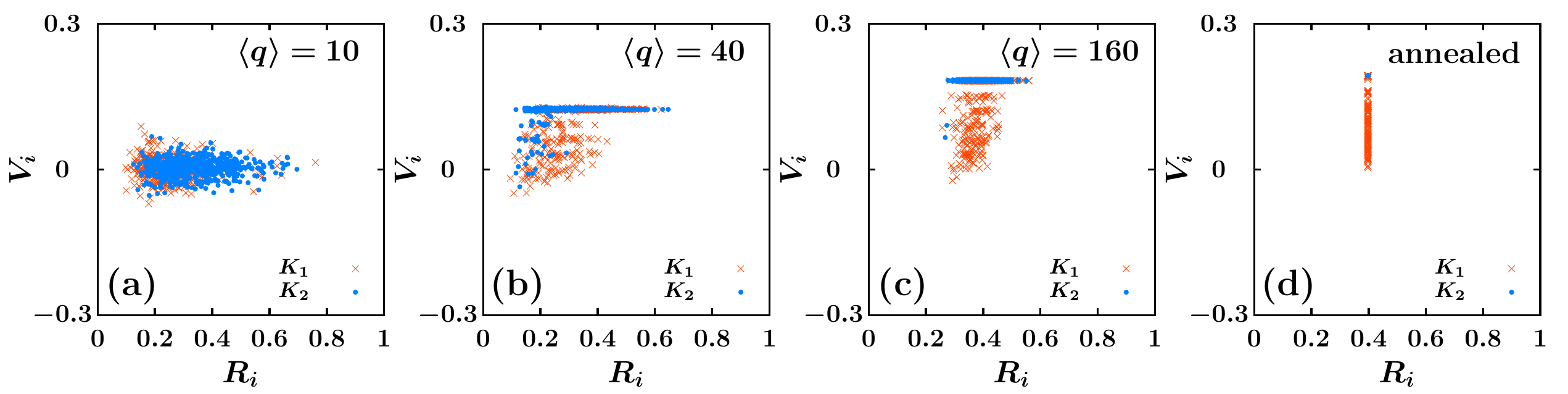}
	\caption{Time-averaged mean velocity $V_i\equiv \frac{1}{T}\int_0^T \dot\theta_i(t) dt$ versus the local order parameter $R_i\equiv\frac{1}{T}\int_0^T \left|\frac{1}{q_i}\sum_j a_{ij}e^{i\theta_j(t)}\right| dt$. Simulation conditions are the same as in Fig.~\ref{fig:fig-correlation}.}
	\label{fig:fig-local-op}
\end{figure}
Finally, we looked at snapshots of the oscillators' phases and phase velocities [Fig.~\ref{fig:fig-phase-scatter}] and the time-averaged mean velocity $V_i\equiv \frac{1}{T}\int_0^T \dot\theta_i(t) dt$ versus local order parameter $R_i\equiv\frac{1}{T}\int_0^T \left|\frac{1}{q_i}\sum_j a_{ij}e^{i\theta_j(t)}\right| dt$ [Fig.~\ref{fig:fig-local-op}]. Note that $\langle q\rangle=10$ has large fluctuations, and it cannot divide the $K_1$ and $K_2$ oscillators into two groups [Fig.~\ref{fig:fig-phase-scatter}(a)]. As the network becomes more connected, the clusters of $K_1$ and $K_2$ oscillators become much clearer, as shown in Figs.~\ref{fig:fig-phase-scatter}(b), (c), and (d). In Fig.~\ref{fig:fig-local-op}(a), the TW regime is absent. In Figs.~\ref{fig:fig-local-op}(b) and (c), the TW order becomes clearer, and the fluctuations in the local order parameter $R_i$ gradually decrease. Note that in the annealed network [Fig.~\ref{fig:fig-local-op}(d)], the local order parameter has become homogeneous on all the oscillators. Hence, for the annealed network, the mean-field theory becomes valid. From Figs.~\ref{fig:fig-correlation} and~\ref{fig:fig-local-op}, we conclude that as the network becomes more connected, the homogeneity of the local environment is eventually restored.

\section{Summary and discussion}\label{sec:summary}
We investigated the synchronization transitions in the $c$-KM with a uniform intrinsic frequency distribution on completely connected networks and on ER networks. For the former case, we presented the SC solutions and the critical exponent of the IC-to-$\pi$ hybrid synchronization transition explicitly. Phase diagrams including the IC, $\pi$, and TW phases were obtained analytically for different ratios $Q\equiv |K_1|/K_2$ of the coupling constants as a function of $p$ and $\gamma$, fractions of nodes with positive coupling constant $K_2$, and half-widths of the uniform distribution, respectively. Various synchronization transition types and critical behaviors were identified analytically or numerically. The linear stability of each SC solution was tested by numerical simulations.

Further, the synchronization transition was investigated on the random network. Some of the synchronization features were found to depend on the network mean degree. When the mean degree is large, the behaviors of the synchronization transition are similar overall to those on the completely connected networks. However, when the mean degree is small, the TW phase was not observed in the numerical simulations, even though the mean-field solution on the annealed ER networks predicts the existence of the TW state when the other control parameters are suitable. Transition features of all-to-all networks and ER networks appear differently. We note that the mean-field theory cannot provide a complete understanding of the competing synchronization transition on the ER networks and that advanced mean-field theories are necessary.

\ack This work was supported by the National Research Foundation of Korea by Grant No. NRF-2014R1A3A2069005.

\end{document}